\documentclass{elsart}

\usepackage{graphicx}

\usepackage{lineno}

\usepackage{amssymb}

\begin{document}


\newcommand{\tritium}{$^3{\rm H}$}
\newcommand{\ttwo}{${\rm T}_2$}
\newcommand{\micron}{$\mu$m}
\newcommand{\isotope}[2]{$^{#2}{\rm #1}$}


\begin{frontmatter}

\title{Performance of a TiN-coated monolithic silicon pin-diode array under mechanical stress}


\author{B.\,A.\,VanDevender\corauthref{brent}\thanksref{PNNLNow}},
\author{L.\,I.\,Bodine},
\author{A.\,W.\,Myers\thanksref{PNNLNow}},
\author{J.\,F.\,Amsbaugh},
\author{M.\,A.\,Howe\thanksref{MarkNow}},
\author{M.\,L.\,Leber\thanksref{MichelleNow}},
\author{R.\,G.\,H.\,Robertson},
\author{K.\,Tolich},
\author{T.\,D.\,Van\,Wechel},
\author{B.\,L.\,Wall}

\address{Center for Experimental Nuclear Physics and Astrophysics, and Department of Physics, University of Washington, Seattle, WA 98195, USA}
\thanks[PNNLNow]{Present Address: Pacific Northwest National Lab, Richland, WA 99354, USA}
\thanks[MarkNow]{Present Address: Department of Physics, University of North Carolina, Chapel Hill, NC 27599, USA}
\thanks[MichelleNow]{Present Address: Department of Physics, University of California, Santa Barbara, CA 93106, USA}
\corauth[brent]{Corresponding author.  brent.vandevender@pnnl.gov}


\begin{abstract}
The Karlsruhe Tritium Neutrino Experiment (KATRIN) will detect tritium beta-decay electrons that pass through its electromagnetic spectrometer with a highly-segmented monolithic silicon $pin$-diode focal-plane detector (FPD).  This $pin$-diode array will be on a single piece of 500-\micron-thick silicon, with contact between titanium nitride (TiN) coated detector pixels and front-end electronics made by spring-loaded pogo pins.  The pogo pins will exert a total force of up to 50\,N on the detector, deforming it and resulting in mechanical stress up to 50\,MPa in the silicon bulk.  We have evaluated a prototype $pin$-diode array with a pogo-pin connection scheme similar to the KATRIN FPD.  We find that pogo pins make good electrical contact to TiN and observe no effects on detector resolution or reverse-bias leakage current which can be attributed to mechanical stress.
\end{abstract}

\begin{keyword}
silicon, $pin$ diode, mechanical stress, leakage current, titanium nitride, pogo pin.
\PACS 29.40.Gx \sep 29.40.Wk
\end{keyword}

\end{frontmatter}

\begin{linenumbers}
\section{Introduction}\label{sec:intro}

The Karlsruhe Tritium Neutrino Experiment (KATRIN) is a direct, model-independent search for the absolute mass of the electron antineutrino~\cite{KATRINDesignReport04}.
The highest energy electrons from the beta decay of molecular tritium (\ttwo) will be selected by a MAC-E spectrometer~\cite{LobashevAndSpivak} and tagged with a focal-plane detector (FPD).  The overall sensitivity to antineutrino mass depends critically on the minimization of backgrounds, which can be achieved by placing the FPD in extreme high vacuum (XHV) ($p \sim 10^{-9}$\,Pa) and avoiding materials with high natural radioactivity near the FPD.  These conditions constrain the design of the FPD and preclude many standard detector construction materials and techniques.  Therefore we have developed a novel scheme to support and read out the FPD.

\begin{figure}
\begin{center}\includegraphics[width=.9\textwidth]{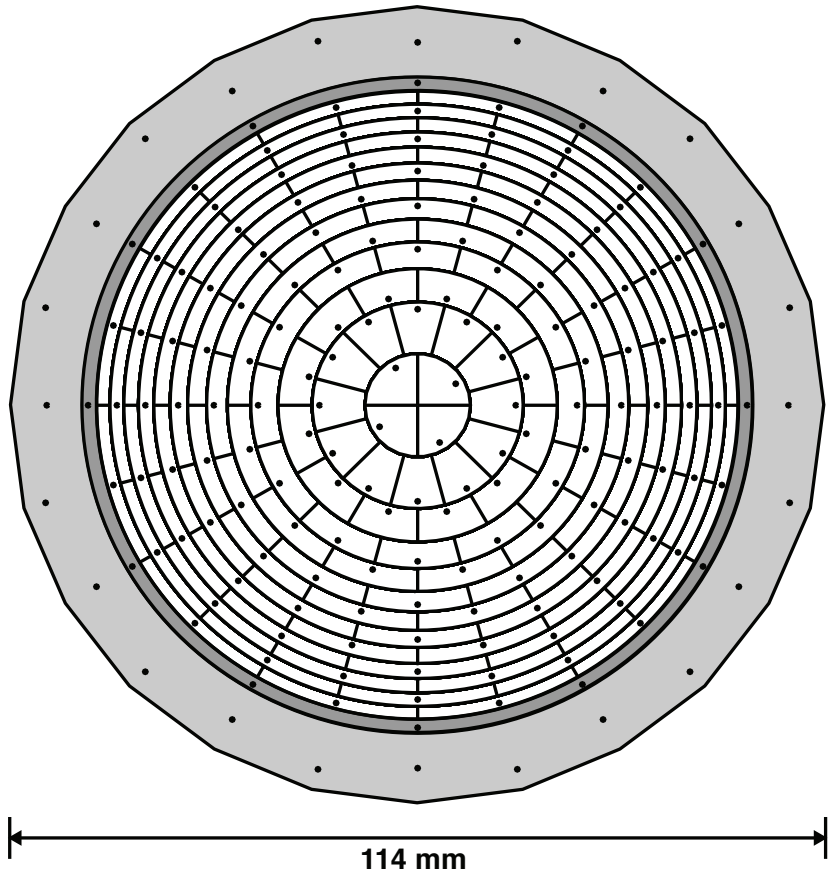}\end{center}
\caption{KATRIN FPD, viewed from the pixelated (back) side.  Lighter gray fill at the edge denotes the bias region which wraps around to the opposite side.  Darker gray fill denotes the guard ring.  White fill denotes individual $pin$-diode  pixels.  Dots denote points on the device that are contacted by pogo pins.  See text for details.}
\label{fig:FPD}
\end{figure}
The KATRIN FPD (Fig.~\ref{fig:FPD}) is a monolithic silicon $pin$-diode array manufactured by Canberra, Belgium.  The $n$-type silicon substrate wafer is 500\,\micron \ thick with a diameter of 114\,mm.  It is bare, with no housing or backing of any kind.  The entrance, or front side, is uniformly $n^{++}$-doped with no segmentation.  The back side is $p^{++}$-doped in 148 individual 44.1-mm$^2$ pixels arranged in a circular dartboard pattern surrounded by a continuous guard ring.  The $\left<111\right>$ crystal orientation is perpendicular to the surface.  The outermost pixels extend to a diameter of 90\,mm.  A guard ring extends from a diameter of 90\,mm out to 94\,mm.  The pixels and guard ring are covered with titanium nitride (TiN) for electrical contact to the front-end electronics.  The guard ring is to be held at signal reference potential to sink surface currents that might otherwise flow between the outer pixels and regions of the surface held at bias potential.  Outside the guard ring there is a coating of TiN (but no doping) from a diameter of 100\,mm extending around the edge of the wafer to the $n^{++}$ doping on the front side.  The TiN on the front side extends a few mm from the edge, but the front is otherwise unmetallized.  This configuration allows the bias potential on the front side to be applied via connections made on the back side adjacent to signal connections.  TiN was chosen as a coating over more commonly used metals such as aluminum because it is non-oxidizing.  The resistance of thin oxide layers is likely to have prevented good electrical contact to the front-end electronics.  Silver, though nonoxidizing, is excluded because it would create a background from an atomic fluorescence near in energy to the highest energy \ttwo \ beta-decay electrons of interest to KATRIN.

The radiopurity requirements also limit the choice of techniques for connecting the detector to its front-end electronics.  In particular, wire bonding is avoided because of the radioactivity of ceramic substrates typically used in the process.  The process of bonding ceramic to silicon can also create localized regions of very high stress because of dissimilar coefficients of thermal expansion.   Instead, the connection between detector and electronics is made by an array of spring-loaded pogo pins.  The pins are manufactured by Interconnect Devices, Inc. (item SS-30 with tip option J).  They are made of inherently radiopure and XHV-compatible materials.  The FPD is supported at a diameter of 99\,mm on its front side while the pins press into the TiN on the back side.  The front-end electronics are on the opposite side of a feedthrough in a region of high vacuum ($p \sim 10^{-5}$\,Pa).

There will be a total of 184 pins pushing on the FPD; one for each of the 148 pixels, 12 contacting the guard ring and 24 applying bias to the outermost ring.  These pins will exert a total force of up to 50\,N on the FPD, which will deform it, causing mechanical stress up to 50\,MPa in the silicon bulk.  Mechanical stress reduces the band gap energy in silicon and thus increases the bulk-generated reverse-bias leakage current through a $p/n$ junction with respect to unstressed silicon~\cite{FuruhashiAndTaniguchi}.   In this model, 50\,MPa of stress in the region of the junction causes approximately 25\% increase in leakage current.  Stress can also induce an increase of approximately 10\% in junction capacitance for $p^+/n$ and $n^+/p$ junctions under a few volts of reverse bias~\cite{Gopinath_etal}.  Both effects would tend to degrade the energy resolution of a silicon $pin$-diode detector by several percent, if they apply to such devices.  Most troubling for KATRIN is a report~\cite{Furgeri_et_al06} demonstrating that stress causes leakage current to increase by up to four orders of magnitude in a few specimens of outer silicon tracker sensors at the Large Hadron Collider's Compact Muon Solenoid (CMS).  No mechanism for this correlation is proposed, though bulk defects in the silicon are ruled out as the cause.  The behavior was observed only in a small subset of the production line CMS devices from one of two manufacturers.  Prototypes from the same manufacturer did not show the correlation.  The mechanisms described in~\cite{FuruhashiAndTaniguchi} can not account for such large increases.  Order-of-magnitude leakage-current increases are predicted for stressed small-pitch devices~\cite{Smeys_etal}, though it seems unlikely that that model would apply to some but not all of the CMS devices.

It is not clear whether the expected mechanical stress would cause similar behavior in the KATRIN FPD.  In this work, we demonstrate the efficacy of the KATRIN FPD connection scheme by implementing it on a prototype $pin$-diode array similar to the FPD in dimensions and doping architecture.  Properties of this detector are measured while unstressed and for a range of pogo-pin applied stresses up to a maximum of 30--42\,MPa.  No stress-induced effects on the detector performance are observed.

\section{Detector and Apparatus}

\begin{figure}
\begin{center}\includegraphics[width=.9\textwidth]{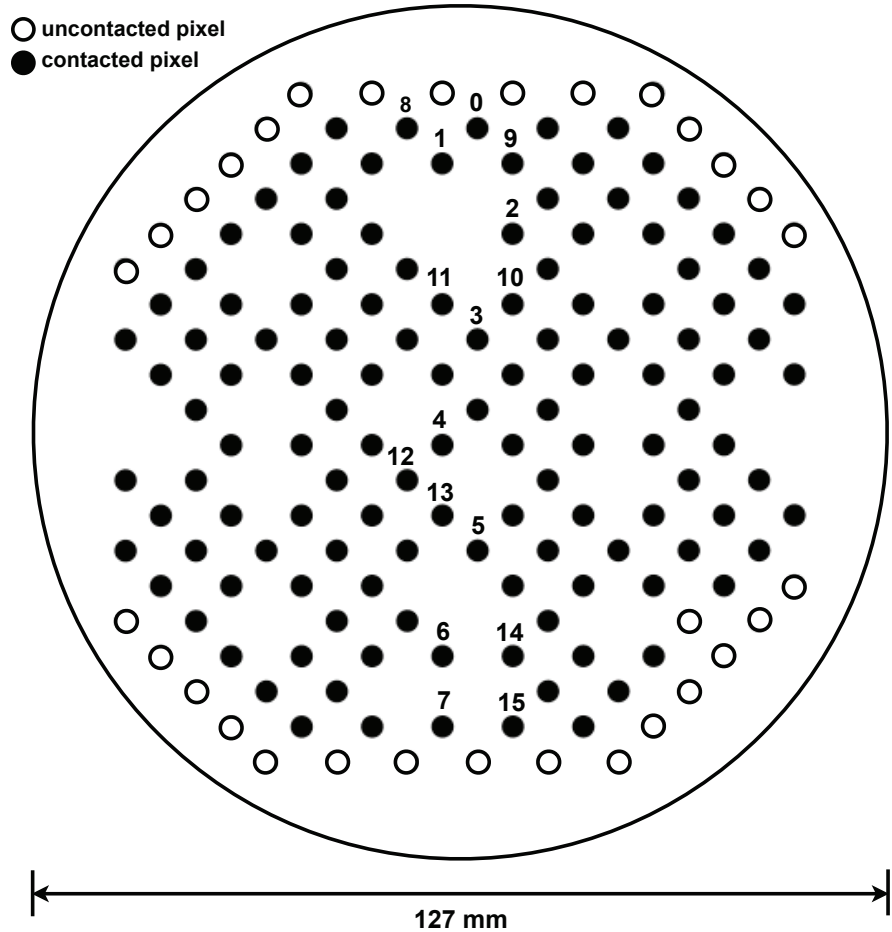}\end{center}
\caption{Pixel layout of the prototype $pin$-diode array used for these measurements, viewed from the pixelated side.  Solid circles are TiN-coated actively doped $pin$-diode regions which are contacted by pogo pins.  Open circles are TiN-coated actively doped $pin$-diode regions which are not contacted by pogo pins.  All pogo pins contact at the center of a diode region.  The numbers 0--15 appear just above the sixteen channels instrumented for these measurements.  See text for details.}
\label{fig:CARACMapBW}
\end{figure}
Canberra, Belgium provided a prototype $pin$-diode array for evaluation (Fig.~\ref{fig:CARACMapBW}).  The prototype's doping architecture is identical to that of the FPD except that the prototype has a thicker $n^{++}$ layer on the entrance side and its window is metalized with aluminum.   Like the FPD, the prototype array is monolithic on a 500-\micron-thick silicon wafer with the $\left<111\right>$ crystal orientation perpendicular to the surface.  The back side is also coated with TiN for ohmic contact to the pixels.  The wafer measures 127\,mm in diameter and has 158 8-mm$^2$ pixels.
The apparatus constructed for these tests allows for measurements of the prototype's basic $pin$-diode detector properties under controlled applied loads from a pogo-pin array.  The pogo-pin array is lowered from above.  The entrance side of the detector faces down towards a sealed 10-$\mu$Ci \isotope{Am}{241} source, placed 6.4\,cm below the detector.  Preamplifiers are mounted above the pogo-pin array.  The entire apparatus resides inside an electrically-shielded light-tight enclosure.

\begin{figure}
\begin{center}\includegraphics[width=.9\textwidth, angle=270]{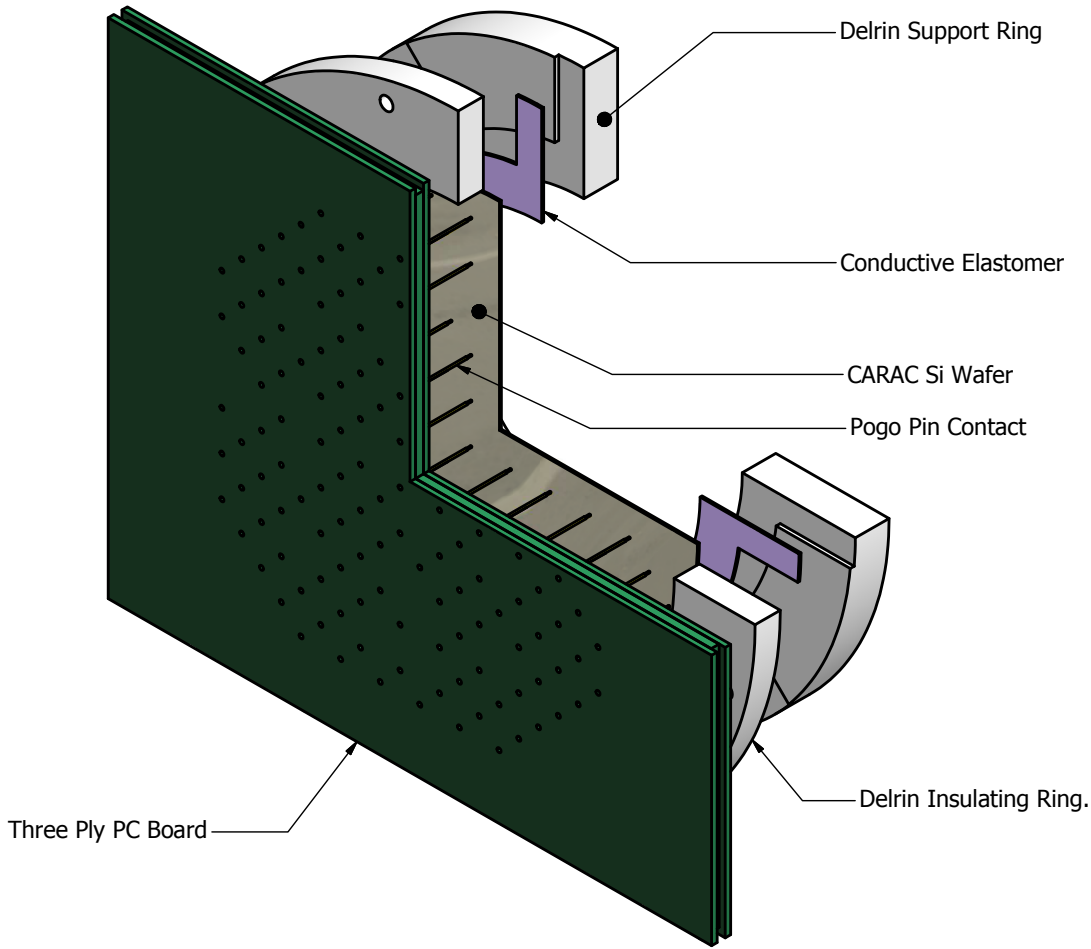}\end{center}
\caption{An exploded cut-away view of the prototype apparatus.}
\label{fig:apparatus}
\end{figure}

The detector is held firmly between two Delrin rings with inner diameters of 110\,mm.  A ring of electrically conductive elastomer sits between the lower ring and the detector in a recess cut into the Delrin.  The recess is slightly shallower than the thickness of the uncompressed elastomer so that when the upper ring is attached the elastomer is compressed for good electrical contact to the detector.  The recess extends to the outer edge of the holder in eight channels.  Positive bias potential is applied to the detector via wires pinched between the elastomer and the upper Delrin ring in each of these channels.  For control measurements of the detector's performance with no stress, an additional plug of Delrin and elastomer is inserted to support the detector from below the front side.  The apparatus is shown in Fig.~\ref{fig:apparatus}.

Electrical contact to each pixel is made by a spring-loaded conducting pogo pin.  The pins are manufactured by Interconnect Devices, Inc.\ (item number SS30  with a rounded tip [option J] and a stainless steel spring).  Each pin is preloaded such that it takes a force of 0.137\,N to initiate any compression of the plunger and an additional force of 0.133\,N per 1000\,\micron \ of compression.

The pogo-pin array is aligned with the detector such that each pin contacts the center of one pixel.  Close mechanical tolerances between the array and the detector holder preclude contacting 30 of the outermost pixels, and one interior pin was damaged during construction of the apparatus, leaving a total of 127 pogo pins applying force to the detector.  The pins are mounted and soldered into a stack of five 1.57-mm-thick FR4 circuit boards which is much more rigid than the detector.  The pins are positioned such that the variation in the relaxed vertical position of the tips is less than 25\,\micron.   A micrometer sets the pogo-pin array translation with an accuracy of 25\,\micron, up to a maximum of 1020\,\micron \ with respect to initial detector contact.

\begin{figure}
\begin{center}\includegraphics[width=.7\textwidth]{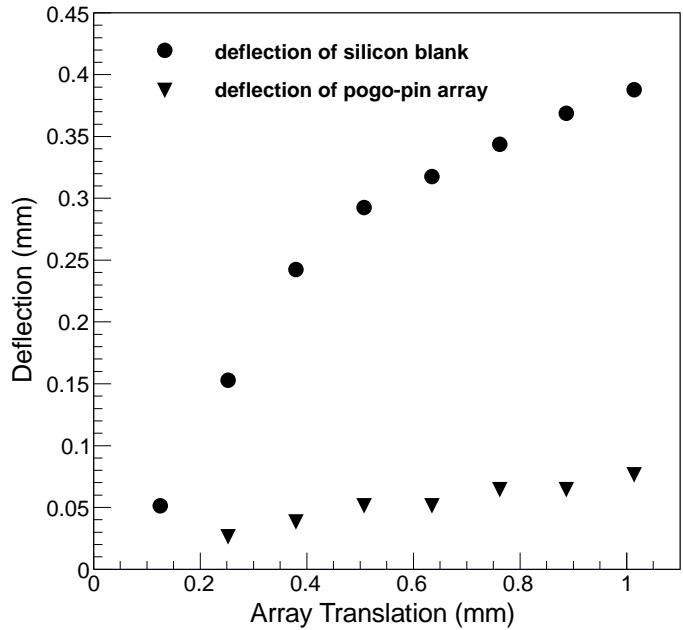}\end{center}
\caption{Mechanical performance of the apparatus with a blank silicon wafer installed.  The abscissa is the distance which the base of the pogo-pin array has translated since initial contact with the detector.}
\label{fig:deflections}
\end{figure}

Uniformity of the applied load is ensured through precise alignment of the pogo-pin array relative to the detector.  The height of all surfaces nominally parallel to the detector deviates less than 25\,\micron, even under the maximum loads allowed by the apparatus.  A blank silicon wafer with identical dimensions as the prototype detector was installed into the apparatus.  Fig.~\ref{fig:deflections} shows the displacement of the center of the blank wafer and the smaller deflection of the center of the pogo-pin array caused by the reaction force.
Due to these deflections, care must be taken to ensure that all pins are compressed at least to the manufacturer's specification of 380\,\micron \ because electrical contact is inconsistent when the pins are under-compressed.  The array must be translated more than 760\,\micron \ past initial contact with the detector so that pins near the center of the detector will meet the minimum requirement.  As can be seen in Fig.~\ref{fig:deflections}, when the array is translated 760\,\micron, the pin compression in the center is reduced by the 340\,\micron \ deflection of the detector in addition to the 60\,\micron \ deflection of the array itself, resulting in only 360\,\micron \ compression of pins at the center.

\section{Forces and Stress}

Calculation of the total force on the detector by the entire array is complicated by the deformation of the detector under the applied load.  The compression of each pin and deflection of the detector depend on the radial position in a way that is not exactly known.  The more relevant quantity is the maximum stress in the detector, which can be estimated without knowing the total load, provided that the displacement $d$ of the center of the detector is known (Fig.~\ref{fig:deflections}).  Mechanically, our detector is a circular plate with fixed edges.  The appropriate relations are~\cite{MachineryHandbook}:
\begin{equation}\label{eqn:stress1}
S = \frac{0.24W}{t^2} \ {\rm and} \ d = \frac{0.0543WR^2}{Et^3},
\end{equation}
where $S$ is the maximum stress in the silicon, $W$ is the total load, $t$ is the thickness of the detector, $R$ is the radius of the supporting ring, and $E$ is the Young's modulus of silicon.  The load, $W$, is trivially eliminated:
\begin{equation}\label{eqn:stress2}
S = \frac{0.24}{0.0543} \frac{dEt}{R^2}.
\end{equation}
The Young's modulus of silicon ranges from 130--185\,GPa, depending on the crystal orientation with respect to the direction of applied stress.  These formulae are appropriate for uniformly distributed loads and deflections of less than 50\% of the plate thickness.  However, the load on our detector has some position dependence due to the variation in pogo-pin compression caused by bowing of the detector, and our detector bows by 80\% of its thickness under the maximum load allowed by the apparatus (Fig.~\ref{fig:deflections}).  Therefore we do not expect stresses calculated with equations (\ref{eqn:stress1}) and (\ref{eqn:stress2}) to be very precise, but take them to be good estimates of the lower limit with the correct order of magnitude.  With these caveats in mind, equation (\ref{eqn:stress2}) indicates that our detector will have a maximum stress of 30--42\,MPa at the maximum array translation allowed by our apparatus. The stated range is due to the unknown orientation of local stresses within the detector bulk.

\section{Electronics}

The custom-made charge-sensitive preamplifiers are prototypes of those to be used for KATRIN.  Because of limited availability of preamplifiers, only eight pixels of the detector are tested simultaneously.  A total of sixteen pixels were tested in two separate series of measurements.  They were chosen to give a representative sample across a diameter of the detector.  The stress is not uniform throughout the volume, but the circular symmetry of the holder guarantees that any stress depends only on the distance from the center of the detector.  During each pass, all pixels neighboring the eight active channels were held at ground to ensure uniform electric fields in the active pixels and also to provide an alternative path for any surface currents which might flow on the detector.  The ground connection was temporarily removed to test for stress-induced surface currents.  No difference was observed.

All of the leakage current through a detector pixel passes through a 5\% precision 500\,M$\Omega$ resistor on its preamplifier, adding a dc offset to the output signal.  We compute the leakage current as the difference between the dc offsets of the output signal when the detector is biased to $V_{\rm bias}$ and when it is floating at $V_{\rm float}$ with the bias supply physically disconnected, divided by the resistance:
\begin{equation}\label{eqn:leakage}
I_{\rm leak}(V_{\rm bias}) = \left[ V_{\rm dc}(V_{\rm bias}) - V_{\rm dc}(V_{\rm float}) \right]/500\,{\rm M}\Omega.
\end{equation}
This relation was validated to an accuracy of $<2$\%  using a picoammeter in series with the single bias supply such that it read the total current drawn by all eight instrumented channels.  The picoammeter introduced a small amount of noise to the output signals, so it was removed during normal running conditions.  The bias potential is $V_{\rm bias} = 100$\,V for all measurements reported here.

\begin{figure}
\begin{center}\includegraphics[width=\textwidth]{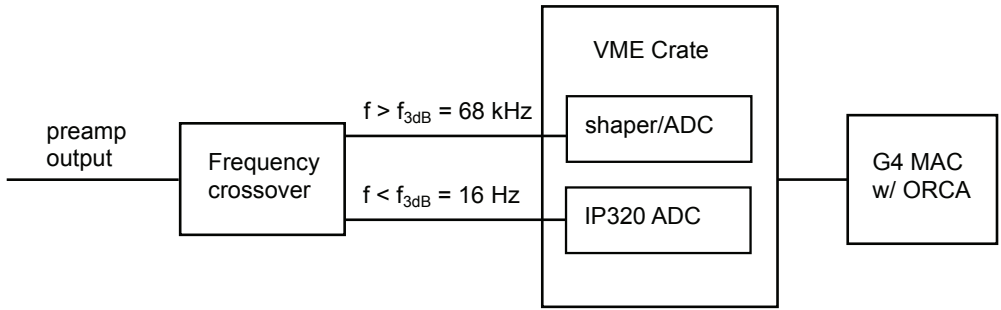}\end{center}
\caption{Signal and data flow.  The preamplifier outputs carry information about the energy of events in high frequency components while the dc component is proportional to the leakage current through the detector.  The crossover segregates these frequency regimes so that they can be read out by appropriate devices.}
\label{fig:signalFlow}
\end{figure}

Detector outputs are split into two branches by a custom-made frequency crossover circuit.  One branch carries components of the signal above a 3-dB cutoff frequency of 68\,kHz into a VME-based shaper/ADC unit described in~\cite{ncdnim}.  The dc offsets, $V_{\rm dc}$, are carried on the other branch into an Acromag IP320A ADC housed in the same VME crate.  Data are acquired from the VME crate by ORCA software~\cite{ORCA}  running on a Macintosh computer.  The setup is shown schematically in Fig.~\ref{fig:signalFlow}.  

The voltages on the dc lines are sampled once per second and these samples are then averaged during two-minute intervals.  The measurements for each translation of the pogo-pin array consist of two such intervals -- one to establish $V_{\rm dc}(V_{\rm float})$ for every channel before the detector is biased and another to establish $V_{\rm dc}(V_{\rm bias})$ for every channel once the detector is at full bias.  The statistical uncertainties in the leakage currents are due to the rms fluctuation of the sampled voltages.  The resistor precisions are not taken into account since they affect all measurements with a given preamplifier by the same proportion, and we are interested only in changes of the leakage currents.  An additional independent systematic uncertainty of $\pm5$\,pA is added to account for the temperature-dependent gate leakage of the FET at the preamplifier input (see below).

\section{Temperature Variations}
The temperature of the apparatus is monitored but it is not controlled.  The temperature inside the light-tight enclosure is measured with a thermistor mounted on the lower surface of the Delrin holder but exposed to air.  The temperatures of the detector and electronics are not directly measured.  The temperature is recorded at the beginning and end of each two-minute dc-offset sampling period.  The nominal temperature for a single leakage current measurement is the average of four readings, two for each of two sampling periods.   The nominal temperature range for the measurements presented here is 24.4--25.6\,$^\circ$C,  though the four individual values which determine any particular nominal temperature never differ by more than 0.1$^\circ$\,C.  We assume that fluctuations in the detector and electronics temperatures are the same size as the fluctuations read on the thermistor.  We therefore take the uncertainty in the nominal temperature for each measurement to be 0.1\,$^\circ$C.

\section{Results}
Our results are demonstrated in two ways.  The first examines the pulse-height resolution of \isotope{Am}{241} gamma-rays as the stress in the detector is increased.  The second method is a direct measurement of leakage currents as discussed above.  Neither method reveals any effects attributable to the stress.  The results are qualitatively the same for all tested pixels.  For brevity, we present results from pixel 8 only.

\begin{figure}
\begin{center}\includegraphics[width=.7\textwidth]{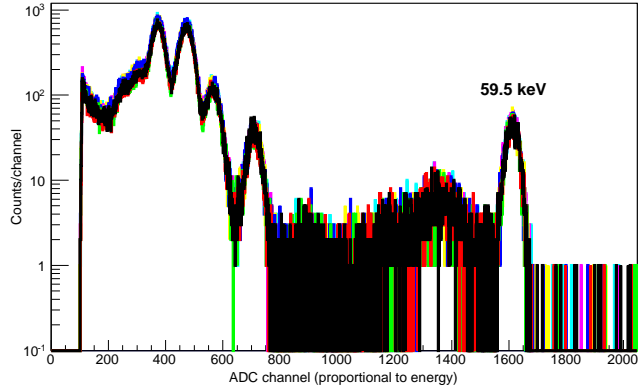}\end{center}
\caption{The pulse-height spectra of \isotope{Am}{241} gamma rays from a single  typical pixel for seven different translations of the pogo-pin array in steps of 130\,\micron \ increasing from the initial compression of 380\,\micron.  Different colors represent different translations.  There is no statistically significant difference between these spectra.}
\label{fig:spectra}
\end{figure}

Fig.~\ref{fig:spectra} shows the raw \isotope{Am}{241} gamma-ray pulse-height spectra obtained with seven different pogo-pin array translations from a typical pixel of our detector.
The peak corresponding to the 59.5\,keV gamma is labelled for reference.  The recognizable gamma-ray spectrum of \isotope{Am}{241} demonstrates the basic functionality of our mounting scheme -- the detector works under stress and the pogo pins make good electrical contact to the TiN-coated pixels.  

\begin{figure}
\begin{center}\includegraphics[height=.25\textheight]{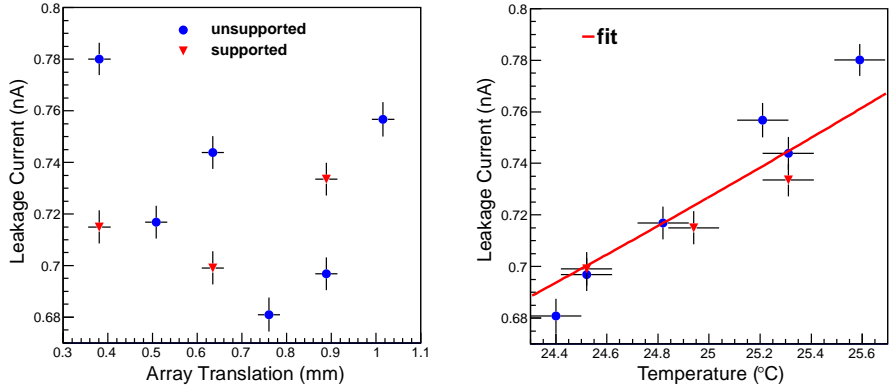}\end{center}
\caption{Leakage current measurements from a typical pixel.  Circles represent measurements taken on an unsupported detector.  Triangles represent control measurements with the support in place.  The left plot shows the leakage current versus the array translation.  The right plot shows the same leakage current measurements versus the temperature at the time of the measurement.  The data fit well with a function of the form of equation (\ref{eqn:IvsT}).}
\label{fig:rawLeakage}
\end{figure}

Fig.~\ref{fig:rawLeakage} shows the leakage current measurements from a typical pixel.
The left plot shows the leakage current versus the array translation, which corresponds to increasing mechanical stress from left to right.  The right plot shows the same leakage currents versus the temperature at the time they were measured.  Leakage currents in diode detectors are thermal in nature so it is necessary to account for temperature dependence when searching for stress-induced changes.  The thermally generated leakage current $I$ is proportional to the rate of electron-hole pair creation and is thus related to the absolute temperature $T$ by~\cite{Knoll} 
\begin{equation}\label{eqn:IvsT}
I(T) = N T^{3/2} e^{-E_g/2 k_{\rm B} T},
\end{equation}
where $E_g$ is the bandgap energy and $k_{\rm B}$ is the Boltzmann constant.  The normalization $N$ accounts for the device geometry and all factors contributing to charge-collection efficiency in the electronics (recombination etc.).  The current measurements made with the detector unsupported are combined with those made with the detector supported and fit with a single function of the form (\ref{eqn:IvsT}), leaving the normalization $N$ as the only free parameter.  The data fit well ($p(\chi^2, \nu) = 0.07$), indicating that all observed variation in the current measurements can likely be attributed to the temperature dependence of bulk-generated leakage currents.

\section{Conclusions}

We have demonstrated the efficacy of a new mounting scheme for monolithic silicon $pin$-diode arrays.   The scheme consists of contacting TiN coated pixels with an array of spring-loaded pins.  We observed no measurable effects of the associated mechanical stress on the performance of our monolithic silicon $pin$-diode array for bulk stresses up to a few tens of MPa.  The large leakage current increases reported in reference~\cite{Furgeri_et_al06} were not observed in the sixteen tested pixels of our prototype detector.  All observed effects can be attributed to temperature variations.    The pogo-pin contact scheme to be used by KATRIN to connect its FPD to front-end electronics is not expected to have any problems related to the pin forces on the detector.

\section{Acknowledgments}

This research was supported by the US Dept. of Energy Division of Nuclear Physics through grant DE-FG02-97ER41020.  The authors wish to thank Marijke Keters, Mathieu Morelle and the rest of their team at Canberra in Olen, Belgium for technical support of this research and for the design and fabrication of KATRIN FPDs.

\end{linenumbers}


\end{document}